\documentclass[11pt]{article}
\usepackage{moriond,epsfig}

\def\PRD{{\em Phys. Rev.} D}

\newcommand{\be}{\begin{equation}} \newcommand{\ee}{\end{equation}}
\newcommand{\bea}{\begin{eqnarray}} \newcommand{\eea}{\end{eqnarray}}
\newcommand{\el}{\nonumber \\}
\newcommand{\re}[1]{(\ref{#1})}
\newcommand{\pat}{\partial}

\newcommand{\bx}{\bar{x}}

\newcommand{\CQG}[1]{Class. Quant. Grav. {\bf #1}}
\newcommand{\GRG}[1]{{\it Gen. Rel. Grav.} {\bf #1}}

\begin{document}
\vspace*{4cm}
\title{BACKREACTION OF LINEAR PERTURBATIONS AND DARK ENERGY\footnote{Contribution to the proceedings of the 2004 Moriond workshop "Exploring the Universe", La Thuile, Italy, March 28-April 4. Based on \cite{Rasanen:2003}, where more details and references can be found.}}

\author{SYKSY R\"{A}S\"{A}NEN}

\address{Theoretical Physics, University of Oxford, 1 Keble Road,\\
Oxford OX1 3NP, UK}

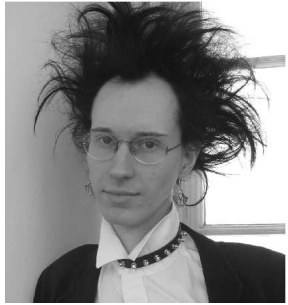
\maketitle\abstracts{A simple discussion on the
backreaction of inhomogeneities in cosmology, focusing 
on the possibility that it could explain the present
acceleration and solve the coincidence problem.}

\section{Introduction}

\paragraph{The coincidence problem.}

Perhaps the most surprising observation in modern cosmology
is that the expansion of the universe has apparently started
to accelerate in the recent past, at a redshift of probably
less than one. This phenomenon has usually been interpreted
in the context of a homogeneous and isotropic model of the
universe. The acceleration can then be accommodated only by
modifying gravity or by introducing an energy component with
negative pressure (or energy density), 'dark energy', the
magnitude of which is fit to the observations.

These models suffer from the \emph{coincidence problem:}
they do not explain why the energy densities of matter and
dark energy have become comparable only recently. If there
is a dynamical explanation for the acceleration
having started in the near past --instead of it being a
coincidence-- then presumably it is related
to the dynamics we see in the universe. The most significant
change in the universe at small redshifts is the formation of
large scale structure, so it seems a natural possibility that
the observed deviation from the natural prediction of the
homogeneous and isotropic model could be related
to the growth of inhomogeneities in the universe.

\paragraph{The fitting problem.}

The reasoning behind using a homogeneous and isotropic model
is that the universe appears to be homogeneous and isotropic
when averaged over sufficiently large scales. In the
usual approach one \emph{first} takes the average of the metric and
the energy-momentum tensor, and then plugs these smooth quantities
into the Einstein equation. However, physically one should
first plug the inhomogeneous quantities into the Einstein
equation and \emph{then} take the average. Because the Einstein
equation is non-linear, these two procedures are not equivalent:
\bea \label{nc}
  \langle G_{\mu\nu}(g_{\alpha\beta})\rangle \neq G_{\mu\nu}(\langle g_{\alpha\beta}\rangle ) \ ,
\eea

\noindent where $G_{\mu\nu}$ is the Einstein tensor,
$g_{\alpha\beta}$ is the metric and $\langle \rangle $ stands for
averaging. The content of \re{nc}
is that \emph{the average behaviour of an inhomogeneous spacetime
is not the same as the behaviour of the corresponding smooth
spacetime.} Here ``corresponding'' means that the smooth
and average quantities have the same
initial conditions. In other words, the
average properties of an inhomogeneous spacetime (energy
density, expansion rate, \ldots) \mbox{do not
satisfy the Einstein equation.}

This is the \emph{fitting problem} discussed by George
Ellis in 1983 \cite{Ellis}. Simply put, how does one find the
average model which best fits the real inhomogeneous universe?
The difference between the behaviour of average and smooth
quantities is also known as \emph{backreaction}.

The equations satisfied by the average quantities in a
general inhomogeneous spacetime have been obtained
\cite{Buchert}. However, these equations are not closed, which
simply means that different
spacetimes sharing the same initial average values
evolve differently even as far as their averages
are concerned -- not a surprising result.
We have taken (following \cite{Geshnizjani:2002}) a more
modest approach, in which one assumes a
smooth background and studies the effect of perturbations.

\section{The backreaction calculation}

\paragraph{The expansion rate.}

We consider a homogeneous and isotropic universe that is spatially flat
and filled with a pressureless fluid. We assume that
the fluid has scale-invariant adiabatic perturbations
(with zero mean) and an amplitude given
by the measurements of the CMB. In first order
perturbation theory, the metric in the longitudinal gauge reads
\bea \label{metric}
  ds^2 &=& -(1+2\Phi(\bx)) dt^2 + (1-2\Phi(\bx))\, a(t)^2 d\bx^2 \ ,
\eea

\noindent where $a=(t/t_0)^{2/3}$ and $\Phi$ is constant in time.

We want to find out the effect of the perturbations on the
expansion rate measured by a comoving observer.
Since the spacetime is inhomogeneous, the expansion
rate is different for observers in different points,
and to make the comparison to a
homogeneous and isotropic universe, we should take an average.
The expansion rate is a covariantly defined scalar quantity, so
this operation is well defined, but the result depends
on which hypersurface one chooses to take the average on
\cite{Geshnizjani:2002,Unruh:1998}.
We \emph{define} the average expansion rate to be the
average of the expansion rates measured by comoving observers
(weighed by the volume element) whose clocks show
the same proper time $\tau$ (measured from the big bang).
In other words, we average on the hypersurface of
constant proper time $\tau$ of comoving observers.

The covariant definition of the local expansion rate is
$\theta = u^{\mu}_{\ \, ;\mu}$, where
$u^{\mu}$ is the four-velocity of the matter fluid.
From the metric \re{metric}, we find $u^{\mu}$
and from that we obtain the expansion rate $\theta$
and the proper time $\tau$. Then we simply express
$\theta$ in terms of $\tau$ and take the average. Note
that this calculation involves no dynamics: we just take
a given metric and calculate the observable of interest
for that spacetime.

Naively, one might expect the average expansion rate to be
\bea \label{naive}
  \langle \theta\rangle  &\simeq& 3 H_{\tau} \left( 1 + \alpha_1 \langle \Phi\rangle  + \alpha_2 \langle\Phi\rangle^2 + \alpha_3 \langle \Phi^2\rangle  \right) \ ,
\eea

\noindent where $H_{\tau}=2/(3\tau)$ is the background expansion
rate in terms of the proper time, and $\alpha_i$ are some
constants of order one. Here and in what follows we expand only
to second order in $\Phi$ (note that we use results
from first-order perturbation theory, so that this is not a consistent
second order calculation). Since $\Phi$ has zero mean, the
linear term and its square vanish, and since
the amplitude of $\Phi$ is $10^{-5}$, the quadratic term
is of the order of $10^{-10}$ -- completely negligible.

When one actually calculates the expansion rate, the result is not
\re{naive}, but instead
\bea \label{real}
  \langle \theta\rangle &\simeq& 3 H_{\tau} \left( 1 + \beta_1\frac{1}{(a_{\tau} H_{\tau})^2}\langle\nabla^2\Phi\rangle + \beta_2\frac{1}{(a_{\tau} H_{\tau})^4}\langle\nabla^2\Phi\rangle^2 + \beta_3 \frac{1}{(a_{\tau} H_{\tau})^2}\langle\pat_i\Phi\pat_i\Phi\rangle \right. \el
  && \left. + \beta_4 \frac{1}{(a_{\tau} H_{\tau})^2} \langle\pat_i (\Phi\pat_i\Phi)\rangle + \beta_5 \frac{1}{(a_{\tau} H_{\tau})^4} \langle\pat_i (\nabla^2\Phi\pat_i\Phi)\rangle \right) \el
  &\equiv& 3 H_{\tau} \left( 1 + \lambda_1 a_{\tau} + \lambda_2 a_{\tau}^2 \right) \ ,
\eea

\noindent where $\beta_i$ are given constants of order one,
$a_{\tau}=(\tau/\tau_0)^{2/3}$, and $\lambda_i$ are constants.

The plain powers of $\Phi$ that one would expect do not
appear, and instead the backreaction is given by gradients
of $\Phi$. The structure of the backreaction terms in
\re{real} is easy to understand: each gradient has to be
accompanied by $1/a$ for reasons of covariance and by $1/H$
for reasons of dimensionality. Since
$\theta$ is a scalar, gradients have to appear in pairs,
and since the Einstein equation is second order, there are
at most two gradients for each $\Phi$. Since
$1/(a H)^2\propto a$, the backreaction grows
relative to the background, and thus behaves like dark energy.

\paragraph{Backreaction as dark energy.}

For a homogeneous and isotropic spatially flat universe,
the expansion law is (in this case $\theta=3 H$)
\bea \label{thetaFRW}
  \left( \theta/3 \right)^2 &=& \frac{1}{3 M_{Pl}^2}\rho \ ,
\eea

\noindent where $M_{Pl}$ is the Planck mass and $\rho$ is
the energy density, which for matter behaves like $\rho\propto a^{-3}$.

In contrast to \re{thetaFRW}, the average of the inhomogeneous
expansion law is, from \re{real},
\bea \label{thetareal}
  \langle \theta/3\rangle^2 &\simeq& H_{\tau}^2 ( 1 + \lambda_1 a_{\tau} + \lambda_2 a_{\tau}^2 )^2 \el
  &\propto& a_{\tau}^{-3} + 2\lambda_1 a_{\tau}^{-2} + (\lambda_1^2 + 2\lambda_2) a_{\tau}^{-1} + 2\lambda_1\lambda_2 + \lambda_2^2 a_{\tau} \ ,
\eea

\noindent where in the second equality we have taken into account
$H_{\tau}^2\propto a_{\tau}^{-3}$.

An observer in a perturbed universe trying to fit the smooth model
\re{thetaFRW} to the observed expansion rate \re{thetareal}
would conclude that there is a mysterious energy component which is
nowhere to be seen but which affects the expansion rate.
Writing $\rho=\rho_m+\rho_{de}$ where $\rho_m$ is the energy
density of matter, and $\rho_{de}$ is the apparent energy
density of this 'dark energy', one finds from the usual relation
$\rho_{de}\propto a^{-3(1+w)}$ the equations of
state $w=-1/3, -2/3, -1$ and $-4/3$.

The equation of state is negative because the backreaction
grows relative to the background for which $w=0$.
Note that there is nothing unnatural about having an equation
of state which is more negative than $-1$ (as marginally
indicated by observations \cite{Alam:2004,Huterer:2004}).
It does not imply violation of the weak
energy condition, since the 'dark energy' is only a parametrisation
of our ignorance of the real expansion law, and does not correspond
to an actual energy component.

Backreaction has a negative equation of state
that could produce acceleration, so the next question is whether the
effect is significant. In other words: what are the values
of $\lambda_1$ and $\lambda_2$? For perturbations
in the linear regime, we have
\bea
  \label{lambda} \lambda_1 = \beta_1\langle\pat_i\Phi\pat_i\Phi\rangle /H_0^2 \sim 10^{-5} \ \ , \qquad \lambda_2 = \beta_2\langle\pat_i(\nabla^2\Phi\pat_i\Phi)\rangle /H_0^4\sim \langle\delta_0^2\rangle \sim 1 \ ,
\eea

\noindent where $H_0$ is the Hubble parameter today and $\delta_0$
is the density perturbation today. The constant
$\lambda_1$ is easy to evaluate and is negligible, while
$\lambda_2$ is more involved. We have used the relation
$\nabla^2\Phi= -3(a H)^2\delta/2$, valid in the linear regime
of perturbations, which we have taken to end
at $\langle \delta^2\rangle =1$. The quantity to be averaged can
then (by definition) be of order one point by point.
However, since it is a total derivative, the
periodicity implied by the standard decomposition in terms
of Fourier modes puts it artificially to zero. The contribution
of similar terms at order $\Phi^4$

\pagebreak

\noindent could be non-vanishing, since
the square of a total derivative is not a total derivative.
If we would have taken the average of $\theta^2$ instead
of averaging $\theta$ and squaring it, we would have found
\bea \label{theta2}
  \langle(\theta/3)^2\rangle &\simeq& H_{\tau}^2 \left( 1 + \frac{4}{81}\frac{1}{(a H)^4}\langle\nabla^2\Phi\nabla^2\Phi\rangle  \right)
  = H_{\tau}^2 \left( 1 + \frac{1}{9}\langle\delta^2\rangle \right) \ ,
\eea

\noindent where all subdominant terms have been dropped.
For comparison to the observations
it is more correct to take the average first, but \re{theta2} shows
that backreaction from linear perturbations can have a sizeable
effect; taking the cut-off to be at $\langle\delta^2\rangle$=1,
the correction is of the order of 10\%.

The reason that the effect can be large even though the
metric perturbation $\Phi$ is small is that the physics is not
given just by the metric but by its first and second derivatives.
The derivatives of the metric are dimensional quantities,
so they are small or large only with respect to some scale.
In the case of spatial derivatives in cosmology, the
comparison scale turns out to be $a H$, the size of the
horizon, compared to which the spatial gradients are large.

\section{Conclusion}

In summary, the backreaction of linear perturbations on
the expansion rate is non-zero, boosted by powers of
$k^2/(a H)^2$ from the naive expectation of powers of
$\Phi$, and naturally involves a negative
equation of state. For periodic boundary conditions, it
is also numerically negligible at second order in $\Phi$.

In evaluating the backreaction of linear perturbations,
we had to introduce a cut-off to obtain a finite result.
A more realistic treatment would take non-linear perturbations
and structure formation explicitly into account; note that
the presence of $\delta$ relates backreaction directly to the
growth of structure.
This calculation has not been done, but one would expect
the growth of the gravitational potential to increase
the magnitude of backreaction and make its
equation of state more negative. The backreaction of
perturbations breaking away from the linear regime
could thus possibly give 'dark energy' at
the right time to solve the coincidence problem.

Whether or not backreaction turns out to be quantitatively
important, the impact of inhomogeneities on the
expansion rate should be properly evaluated to solve the
fitting problem and make sure that we are using the
right theoretical equations to explain increasingly precise
cosmological observations.

\section*{Acknowledgments}

I thank Jai-chan Hwang for useful discussions at the 
2004 Moriond workshop "Exploring the Universe".
The research has been supported by PPARC grant PPA/G/O/2002/00479,
by a grant from the Magnus Ehrnrooth Foundation and by the
European Union network HPRN-CT-2000-00152,
``Supersymmetry and the Early Universe''.

\end{document}